\documentclass{llncs}
\usepackage{psfig}

\newcommand{\dist}{\hbox{\it dist}}
\newcommand{\sigthin}{\sigma_{\rm thin}}
\newcommand{\sighi}{\sigma_{\rm hi}}
\newcommand{\siglo}{\sigma_{\rm lo}}
\newcommand{\dmax}{d_{\rm max}}

\begin{document}
\bibliographystyle{plain}

\title{Analysis of Approximate Nearest Neighbor Searching
	with Clustered Point Sets\thanks{The support of the National
		Science Foundation under grant CCR--9712379 is gratefully
		acknowledged.}
	}

\author{Songrit Maneewongvatana\inst{1} \and
	David M. Mount\inst{2}}
\institute{
    Department of Computer Science, University of Maryland, College Park,
    Maryland. \email{\tt songrit@cs.umd.edu}.
    \and
    Department of Computer Science and Institute for Advanced Computer
    Studies, University of Maryland, College Park, Maryland.  
    \email{mount@cs.umd.edu}.}

\maketitle
\thispagestyle{empty}

\section{Introduction}\label{intro.sec}

Nearest neighbor searching is the following problem: we are given a set
$S$ of $n$ {\em data points} in a metric space, $X$, and are asked to
preprocess these points so that, given any {\em query point} $q \in X$,
the data point nearest to $q$ can be reported quickly.  Nearest neighbor
searching has applications in many areas, including knowledge discovery
and data mining \cite{fpsu-akddm-96}, pattern recognition and
classification \cite{ch-nnpc-67,dh-pcsa-73}, machine learning
\cite{cs-wnnalsf-93}, data compression \cite{gg-vqsc-92}, multimedia
databases \cite{fsna-qivcqs-95}, document retrieval
\cite{ddflh-ilsa-90}, and statistics \cite{dw-nnmd-82}.

There are many possible choices of the metric space.  Throughout we will
assume that the space is $R^d$, real $d$-dimensional space, where
distances are measured using any Minkowski $L_m$ distance metric.  For
any integer $m \ge 1$, the {\em $L_m$-distance} between points $p =
(p_1,p_2,\ldots,p_d)$ and $q=(q_1,q_2,\ldots,q_d)$ in $R^d$ is defined
to be the $m$-th root of $\sum_{1 \le i \le d} |p_i-q_i|^m$.  The $L_1$,
$L_2$, and $L_{\infty}$ metrics are the well-known Manhattan, Euclidean
and max metrics, respectively.

Our primary focus is on data structures that are stored in main memory.
Since data sets can be large, we limit ourselves to consideration of
data structures whose total space grows linearly with $d$ and $n$.
Among the most popular methods are those based on hierarchical
decompositions of space.  The seminal work in this area was by Friedman,
Bentley, and Finkel \cite{fbf-afbml-77} who showed that $O(n)$ space and
$O(\log n)$ query time are achievable for fixed dimensional spaces in
the expected case for data distributions of bounded density through the
use of kd-trees.  There have been numerous variations on this theme.
However, all known methods suffer from the fact that as dimension
increases, either running time or space increase exponentially with
dimension.

The difficulty of obtaining algorithms that are efficient in the worst
case with respect to both space and query time suggests the alternative
problem of finding {\em approximate} nearest neighbors.  Consider a set
$S$ of data points in $R^d$ and a query point $q \in R^d$.  Given
$\epsilon > 0$, we say that a point $p \in S$ is a {\em
$(1+\epsilon)$-approximate nearest neighbor} of $q$ if
\[
	\dist(p,q) \le (1 + \epsilon)\dist(p^*,q),
\]
where $p^*$ is the true nearest neighbor to $q$.  In other words, $p$
is within relative error $\epsilon$ of the true nearest neighbor.
The approximate nearest neighbor problem has been heavily studied
recently.  Examples include algorithms by Bern \cite{b-acpqhd-93},
Arya and Mount \cite{am-annqf-93}, Arya, et al. \cite{amnsw-oaann-94},
Clarkson \cite{c-aacpq-94}, Chan \cite{c-annqr-97}, Kleinberg
\cite{k-tanns-97}, Indyk and Motwani \cite{im-anntr-98}, and
Kushilevitz, Ostrovsky and Rabani \cite{kor-esann-98}.

In this study we restrict attention to data structures of size $O(dn)$
based on hierarchical spatial decompositions, and the kd-tree in
particular.  In large part this is because of the simplicity and
widespread popularity of this data structure.  A kd-tree is binary tree
based on a hierarchical subdivision of space by splitting hyperplanes
that are orthogonal to the coordinate axes \cite{fbf-afbml-77}.  It is
described further in the next section.  A key issue in the design of the
kd-tree is the choice of the splitting hyperplane.  Friedman, Bentley,
and Finkel proposed a splitting method based on selecting the plane
orthogonal to the median coordinate along which the points have the
greatest spread.  They called the resulting tree an optimized kd-tree,
and henceforth we call the resulting splitting method the {\em standard
splitting method}.  Another common alternative uses the shape of the
cell, rather than the distribution of the data points.  It splits
each cell through its midpoint by a hyperplane orthogonal to its longest
side.  We call this the {\em midpoint split method}.

A number of other data structures for nearest neighbor searching based
on hierarchical spatial decompositions have been proposed.  Yianilos
introduced the {\em vp-tree} \cite{y-dsann-93}.  Rather than using an
axis-aligned plane to split a node as in kd-tree, it uses a data point,
called the vantage point, as the center of a hypersphere that partitions
the space into two regions.   There has also been quite a bit of
interest from the field of databases.  There are several data structures
for database applications based on $R$-trees and their variants
\cite{bkss-rtera-90,srf-rtdim-87}.  For example, the {\em X-tree}
\cite{bkk-xtish-96} improves the performance of the R$^*$-tree by
avoiding high overlap.  Another example is the SR-tree
\cite{ks-stish-97}.  The {\em TV-tree} \cite{ljf-ttish-94} uses a
different approach to deal with high dimensional spaces.  It reduces
dimensionality by maintaining a number of active dimensions. When all
data points in a node share the same coordinate of an active dimension,
that dimension will be deactivated and the set of active dimensions
shifts.

In this paper we study the performance of two other splitting methods,
and compare them against the kd-tree splitting method.  The first,
called {\em sliding-midpoint}, is a splitting method that was introduced
by Mount and Arya in the ANN library for approximate nearest neighbor
searching \cite{ma-alanns-97}.  This method was introduced into the
library in order to better handle highly clustered data sets.   We know
of no analysis (empirical or theoretical) of this method.  This method
was designed as a simple technique for addressing one of the most
serious flaws in the standard kd-tree splitting method.  The flaw is
that when the data points are highly clustered in low dimensional
subspaces, then the standard kd-tree splitting method may produce highly
elongated cells, and these can lead to slow query times.  This splitting
method starts with a simple midpoint split of the longest side of the
cell, but if this split results in either subcell containing no data
points, it translates (or ``slides'') the splitting plane in the
direction of the points until hitting the first data point.  In
Section~\ref{slmid.sec} we describe this splitting method and analyze
some of its properties.

The second splitting method, called {\em minimum-ambiguity}, is a
query-based technique.  The tree is given not only the data points, but
also a collection of sample query points, called the {\em training
points}.  The algorithm applies a greedy heuristic to build the tree in
an attempt to minimize the expected query time on the training points.
We model query processing as the problem of eliminating data points from
consideration as the possible candidates for the nearest neighbor.
Given a collection of query points, we can model any stage of the
nearest neighbor algorithm as a bipartite graph, called the {\em
candidate graph}, whose vertices correspond to the union of the data
points and the query points, and in which each query point is adjacent
to the subset of data points that might be its nearest neighbor.  The
minimum-ambiguity selects the splitting plane at each stage that
eliminates the maximum number of remaining edges in the candidate graph.
In Section~\ref{minamb.sec} we describe this splitting method in greater
detail.

We implemented these two splitting methods, along with the standard
kd-tree splitting method.  We compared them on a number of synthetically
generated point distributions, which were designed to model
low-dimensional clustering.  We believe this type of clustering is not
uncommon in many application data sets \cite{jd-acd-88}.  We used
synthetic data sets, as opposed to standard benchmarks, so that we could
adjust the strength and dimensionality of the clustering.  Our results
show that these new splitting methods can provide significant
improvements over the standard kd-tree splitting method for data sets
with low-dimensional clustering.  The rest of the paper is organized as
follows.  In the next section we present background information on the
kd-tree and how to perform nearest neighbor searches in this tree.  In
Section~\ref{split.sec} we present the two new splitting methods.  In
Section~\ref{empir.sec} we describe our implementation and present our
empirical results.

\section{Background}\label{backgr.sec}

In this section we describe how kd-trees are used for performing exact
and approximate nearest neighbor searching.  Bentley introduced the
kd-tree as a generalization of the binary search tree in higher
dimensions \cite{b-mbstu-75}.  Each node of the tree is implicitly associated
with a $d$-dimensional rectangle, called its {\em cell}.  The root node
is associated with the bounding rectangle, which encloses all of the
data points.  Each node is also implicitly associated with the subset of
data points that lie within this rectangle.  (Data points lying on the
boundary between two rectangles, may be associated with either.)  If the
number of points associated with a node falls below a given threshold,
called the {\em bucket size}, then this node is a leaf, and these points
are stored with the leaf.  (In our experiments we used a bucket size of
one.) Otherwise, the construction algorithm selects a splitting
hyperplane, which is orthogonal to one of the coordinate axes and passes
through the cell.  There are a number of {\em splitting methods} that may
be used for choosing this hyperplane.  We will discuss these in greater
detail below.  The hyperplane subdivides the associated cell into
two subrectangles, which are then associated with the children of this
node, and the points are subdivided among these children according to
which side of the hyperplane they lie.  Each internal node of the tree
is associated with its splitting hyperplane (which may be given as the
index of the orthogonal axis and a cutting value along this axis).

Friedman, Bentley and Finkel \cite{fbf-afbml-77} present an algorithm to
find the nearest neighbor using the kd-trees.  They introduce the
following splitting method, which we call the {\em standard splitting
method}.  For each internal node, the splitting hyperplane is chosen to
be orthogonal to the axis along which the points have the greatest {\em
spread} (difference of maximum and minimum).  The splitting point is
chosen at the median coordinate, so that the two subsets of data points
have nearly equal sizes.  The resulting tree has $O(n)$ size and $O(\log
n)$ height.  White and Jain \cite{wj-assr-96} proposed an alternative,
called the {\em VAM-split}, with the same basic idea, but the splitting
dimension is chosen to be the one with the maximum variance.

Queries are answered by a simple recursive algorithm.  In the basis
case, when the algorithm arrives at a leaf of the tree, it computes the
distance from the query point to each of the data points associated with
this node.  The smallest such distance is saved.  When arriving at an
internal node, it first determines the side of the associated hyperplane
on which the query point lies.  The query point is necessarily closer to
this child's cell.  The search recursively visits this child.  On
returning from the search, it determines whether the cell associated
with the other child is closer to the query point than the closest point
seen so far.  If so, then this child is also visited recursively.  When
the search returns from the root, the closest point seen is returned.
An important observation is that for each query point, every leaf whose
distance from the query point is less than the nearest neighbor will be
visited by the algorithm.

It is an easy matter to generalize this search algorithm for answering
{\em approximate} nearest neighbor queries.  Let $\epsilon$ denote the
allowed error bound.  In the processing of an internal node, the further
child is visited only if its distance from the query point is less than
the distance to the closest point so far, divided by $(1+\epsilon)$.
Arya et al. \cite{amnsw-oaann-94} show the correctness of this
procedure.  They also show how to generalize the search algorithm for
computing the $k$-closest neighbors, either exactly or approximately.

Arya and Mount \cite{am-afvq-93} proposed a number of improvements to
this basic algorithm.  The first is called {\em incremental distance
calculation}.  This technique can be applied for any Minkowski metric.
In addition to storing the splitting hyperplane, each internal node of
the tree also stores the extents of associated cell projected
orthogonally onto its splitting axis.  The algorithm does not maintain
true distances, but instead (for the Euclidean metric) maintains squared
distances.  When the algorithm arrives at an internal node, it knows the
squared distance from the query point to the associated cell.  They show
that in constant time (independent of dimension) it is possible to use
this information to compute the squared distance to each of the
children's cell.  They also presented a method called {\em priority
search}, which uses a heap to visit the leaves of the tree in increasing
order of distance from the query point, rather than in the recursive
order dictated by the structure of the tree.  Yet another improvement is
a well-known technique from nearest neighbor searching, called {\em
partial distance calculation} \cite{bg-imdeavq-85,s-rnnsk-91}.  When
computing the distance between the query point and a data point, if the
accumulated sum of squared components ever exceeds the squared distance
to the nearest point so far, then the distance computation is
terminated.

One of the important elements of approximate nearest neighbor searching,
which was observed by Arya et al. \cite{amnsw-oaann-94}, is that there
are two important properties of any data structure for approximate
nearest neighbor searching based on spatial decomposition. 

\begin{description}
\item[Balance:] The height of the tree should be $O(\log n)$, where
	$n$ is the number of data points.
\item[Bounded aspect ratio:] The leaf cells of the tree should have
	bounded aspect ratio, meaning that the ratio of the longest
	to shortest side of each leaf cell should be bounded above
	by a constant.
\end{description}

Given these two constraints, they show that approximate nearest neighbor
searching (using priority search) can be performed in $O(\log n)$ time
from a data structure of size $O(dn)$.  The hidden constant factors in
time grow as $O(d/\epsilon)^d$.  Unfortunately, achieving both of these
properties does not always seem to be possible for kd-trees.  This is
particularly true when the point distribution is highly clustered.  Arya
et al. present a somewhat more complex data structure called a {\em
balanced box-decomposition tree}, which does satisfy these properties.
The extra complexity seems to be necessary in order to prove their
theoretical results, and they show empirically that it is important when
data sets are highly clustered in low-dimensional subspaces.  An
interesting practical question is whether there exist methods that
retain the essential simplicity of the kd-tree, while providing
practical efficiency for clustered data distributions (at least in most
instances, if not in the worst case).

Bounded aspect ratio is a sufficient condition for efficiency, but it is
not necessary.  The more precise condition in order for their results to
apply is called the {\em packing constraint} \cite{amnsw-oaann-94}.
Define a {\em ball} of radius $r$ to be the locus of points that are
within distance $r$ of some point in $R^d$ according to the chosen
metric.  The packing constraint says that the number of large cells that
intersect any such ball is bounded.

\begin{description}
\item[Packing Constraint:] The number of leaf cells of size at least
	$s$ that intersect an open ball of radius $r > 0$ is bounded
	above by a function of $r/s$ and $d$, but independent of $n$.
\end{description}

If a tree has cells of bounded aspect ratio, then it satisfies the
packing constraint.  Arya et al., show that if this assumption is
satisfied, then priority search runs in time that is proportional to the
depth of the tree, times the number of cells of maximum side length
$r\epsilon/d$ that intersect a ball of radius $r$.  By the packing
constraint this number of cells depends only on the dimension and
$\epsilon$.  The main shortcoming of the standard splitting method is
that it may result in cells of unbounded aspect ratio.

\section{Splitting Methods}\label{split.sec}

In this section we describe the splitting methods that are considered
in our experiments.  As mentioned in the introduction, we implemented
two splitting methods, in addition to the standard kd-tree splitting
method.  We describe them further in each of the following sections.

\subsection{Sliding-Midpoint}\label{slmid.sec}

\begin{figure}[htbp]
\centerline{\psfig{figure=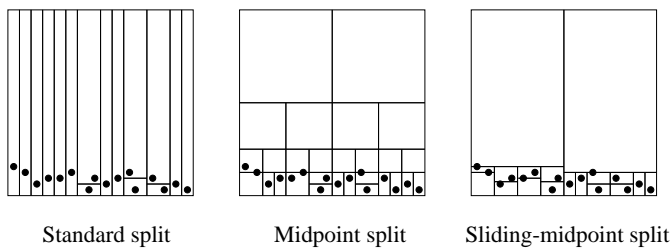,height=1.25in}}
\caption{Splitting methods with clustered point sets.}
\label{slmid.fig}
\end{figure}

The sliding-midpoint splitting method was first introduced in the ANN
library for approximate nearest neighbor searching \cite{ma-alanns-97}.
This method was motivated to remedy the deficiencies of two other
splitting methods, the standard kd-tree splitting method and the
midpoint splitting method.  To understand the problem, suppose that the
data points are highly clustered along a few dimensions but vary greatly
along some the others (see Fig.~\ref{slmid.fig}).  The standard kd-tree
splitting method will repeatedly split along the dimension in which the
data points have the greatest spread, leading to many cells with high
aspect ratio.  A nearest neighbor query near the center of the bounding
square would visit a large number of these cells.  In contrast, the
midpoint splitting method bisects the cell along its longest side,
irrespective of the point distribution.  (If there are ties for the
longest side, then the tie is broken in favor of the dimension along
which the points have the highest spread.)  This method produces cells of
aspect ratio at most 2, but it may produce leaf cells that contain no
data points.  The size of the resulting tree may be very large when the
data distribution is highly clustered data and the dimension is high.

The sliding midpoint method works as follows.  It first attempts to
perform a midpoint split, by the same method described above.  If data
points lie on both sides of the splitting plane then the algorithm acts
exactly as it would for the midpoint split.  However, if a trivial split
were to result (in which all the points lie to one side of the splitting
plane), then it attempts to avoid this by ``sliding'' the splitting plane
towards the points until it encounters the first data point.  More
formally, if the split is performed orthogonal to the $i$th coordinate,
and all the data points have $i$-coordinates that are larger than that
of the splitting plane, then the splitting plane is translated so that
its $i$th coordinate equals the minimum $i$th coordinate among all the
data points.  Let this point be $p_1$.  Then the points are partitioned
with $p_1$ in one part of the partition, and all the other data points
in the other part.  A symmetrical rule is applied if the points all have
$i$th coordinates smaller than the splitting plane.

This method cannot result in any trivial splits, implying that the
resulting tree has size $O(n)$.  Thus it avoids the problem of large
trees, which the midpoint splitting method is susceptible to.  Because
there is no guarantee that the point partition is balanced, the
depth of the resulting tree may exceed $O(\log n)$.  However, based on
our empirical observations, the height of this tree rarely exceeds the
height of the standard kd-tree by more than a small constant factor.

It is possible to generate a cell $C$ of very high aspect ratio, but it
can be shown that if it does, then $C$ is necessarily adjacent to a
sibling cell $C'$ that is fat along the same dimension that $C$ is
skinny.  As a result, it is not possible to generate arbitrarily long
squences of skinny cells, as the standard splitting method could.

The sliding midpoint method can be implemented with little more effort
than the standard kd-tree splitting method.  But, because the depth of
the tree is not necessarily $O(\log n)$, the $O(n\log n)$ construction
time bound does not necessarily hold.  There are more complex algorithms
for constructing the tree that run in $O(n\log n)$ time
\cite{amnsw-oaann-94}.  However, in spite of these shortcomings, we will
see that the sliding-midpoint method, can perform quite well for highly
clustered data sets.

\subsection{Minimum-Ambiguity}\label{minamb.sec}

All of the splitting methods described so far are based solely on
the data points.  This may be quite reasonable in applications where
data points and query points come from the same distribution.  However
this is not always the case.  (For example, a common use of nearest
neighbor searching is in iterative clustering algorithms, such as the
{\em k-means algorithm} \cite{f-camdeic-65,gg-vqsc-92,m-smcamo-67}.
Depending on the starting conditions of the algorithm, the data points
and query points may be quite different from one another.)  If the two
distributions are different, then it is reasonable that preprocessing
should be informed of the expected distribution of the query points, as
well as the data points.  One way to do this is to provide the
preprocessing phase with the data points and a collection of sample
query points, called {\em training points}.  The goal is to compute a
data structure which is efficient, assuming that the query distribution
is well-represented by the training points.  The idea of presenting
a training set of query points is not new.  For example, Clarkson
\cite{c-nnqms-97} described a nearest neighbor algorithm that uses
this concept.

The {\em minimum-ambiguity splitting method} is given a set $S$ of data
points and a training set $T$ of sample query points.  For each query
point $q \in T$, we compute the nearest neighbor of $q$ in $S$ as part
of the preprocessing.  For each such $q$, let $r(q)$ denote the distance
to the nearest point in $S$.  Let $b(q)$ denote the {\em nearest
neighbor ball}, that is, the locus of points (in the current metric)
whose distance from $q$ is at most $r(q)$.  As observed earlier, the
search algorithm visits every leaf cell that overlaps $b(q)$ (and
it may generally visit a large set of leaves).

Given any kd-tree, let $C(q)$ denote the set of leaf cells of the tree
that overlap $b(q)$.  This suggests the following optimization problem,
given point sets $S$ and $T$, determine a hierarchical subdivision of
$S$ of size $O(n)$ such that the {\em total overlap}, $\sum_{q \in T}
|C(q)|$, is minimized.  This is analogous to the packing constraint, but
applied only to the nearest neighbor balls of the training set.  We do
not know how to solve this problem optimally, but we devised the
minimum-ambiguity splitting method as a greedy heuristic.

To motivate our method, we introduce a model for nearest neighbor
searching in terms of a pruning process on a bipartite graph.  Given a
cell (i.e., a $d$-dimensional rectangle) $C$.  Let $S_C$ denote the
subset of data points lying within this cell and let $T_C$ denote the
subset of training points whose such that the nearest neighbor balls
intersects $C$.  Define the {\em candidate graph} for $C$ to be the
bipartite graph on the vertex set $S \cup T$, whose edge set is $S_C
\times T_C$.  Intuitively, each edge $(p,q)$ in this graph reflects the
possibility that data point $p$ is a candidate to be the nearest neighbor
of training point $q$.  Observe that if a cell $C$ intersects $b(q)$ and
contains $k$ data points, then $q$ has degree $k$ in the candidate graph
for $C$.  Since it is our goal to minimize the number of leaf nodes that
overlap $C$, and assuming that each leaf node contains at least one
data point, then a reasonable heuristic for minimizing the number of
overlapping leaf cells is to minimize the average degree of vertices
in the candidate graph.  This is equivalent to minimizing the total
number of edges in the graph.  This method is similar to techniques
used in the design of linear classifiers based on impurity functions
\cite{bfos-crt-84}.

\begin{figure}[htbp]
\centerline{\psfig{figure=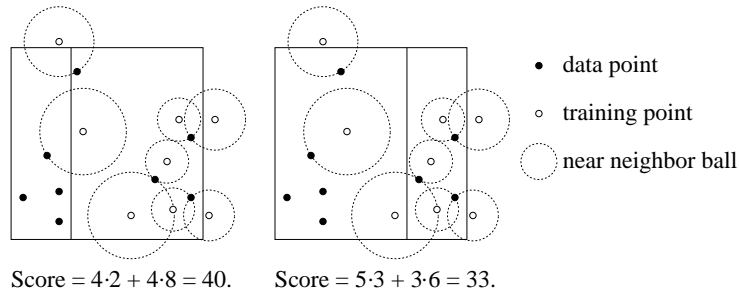,height=1.5in}}
\caption{Minimum ambiguity splitting method.}
\label{minamb.fig}
\end{figure}

Here is how the minimum-ambiguity method selects the splitting
hyperplane.  If $|S_C| \le 1$, then from our desire to generate a tree
of size $O(n)$, we will not subdivide this cell any further.  Otherwise,
let $H$ be some orthogonal hyperplane that cuts $C$ into subcells $C_1$
and $C_2$.  Let $S_1$ and $S_2$ be the resulting partition of data
points into these respective subcells, and let $T_1$ and $T_2$ denote
the subsets of training points whose nearest neighbor balls intersect
$C_1$ and $C_2$, respectively.  Notice that these subsets are not
necessarily disjoint.  We assign a {\em score} to each such hyperplane
$H$, which is equal to the sum of the number of edges in the ambiguity
graphs of $C_1$ and $C_2$.  In particular,
\[
	\hbox{Score}(H) = |S_1| \cdot |T_1| + |S_2| \cdot |T_2|.
\]
Intuitively a small score is good, because it means that the average
ambiguity in the choice of nearest neighbors is small.  The
minimum-ambiguity splitting method selects the orthogonal hyperplane $H$
that produces a nontrivial partition of the data points and has the
smallest score.  For example, in Fig.~\ref{minamb.fig} on the left, we
show the score of the standard kd-tree splitting method.  However,
because of the higher concentration of training points on the right side
of the cell, the splitting plane shown on the right actually has a lower
score, and hence is preferred by the minimum-ambiguity method.  In this
way the minimum-ambiguity method tailors the structure of the tree to
the distribution of the training points.

The minimum-ambiguity split is computed as follows.  At each stage it is
given the current cell $C$, and the subsets $S_C$ and $T_C$.  For each
coordinate axis, it projects the points of $S_C$ and the extreme
coordinates of the balls $b(q)$ for each $q \in T_C$ orthogonally onto
this axis.  It then sweeps through this set of projections, from the
leftmost to the rightmost data point projection, updating the score as
it goes.  It selects the hyperplane with the minimum score.  If there
are ties for the smallest score, then it selects the hyperplane that
most evenly partitions the data points.

\section{Empirical Results}\label{empir.sec}

We implemented a kd-tree in C++ using the three splitting methods: the
standard method, sliding-midpoint, and minimum-ambiguity.  For each
splitting method we generated a number data point sets, query point
sets, and (for minimum-ambiguity) training point sets.  The tree
structure was based on the same basic tree structure used in ANN
\cite{ma-alanns-97}.   The experiments were run on a Sparc Ultra,
running Solaris 5.5, and the program was compiled by the g++ compiler.
We measured a number of statistics for the tree, including its size,
depth, and the average aspect ratio of its cells.

Queries were answered using priority search.  For each group of queries
we computed a number of statistics including CPU time, number of nodes
visited in the tree, number of floating-point operations, number of
distance calculations, and number of coordinate accesses.  In our plots we
show only the number of nodes in the tree visited during the search.  We
chose this parameter because it is a machine-independent quantity, and
was closely correlated with CPU time.  In most of our experiments,
nearest neighbors were computed approximately.

For each experiment we fixed the number of data points, the
dimension, the data-point distribution, and the error bound
$\epsilon$.  In the case of the minimum-ambiguity method, the query
distribution is also fixed, and some number of training points were
generated.  Then a kd-tree was generated by applying the appropriate
splitting method.  For the standard and sliding-midpoint methods the
tree construction does not depend on $\epsilon$, implying that the same
tree may be used for different error bounds.  For the minimum-ambiguity
tree, the error bound was used in computing the tree.  In particular,
the nearest neighbors of each of the training points was computed only
approximately.  Furthermore, the nearest neighbor balls $b(q)$ for each
training point $q$ were shrunken in size by dividing their radius by the
factor $1+\epsilon$.  This is because this is the size of the ball that
is used in the search algorithm.

For each tree generated, we generated some number of query points.
The query-point distribution was not always the same as the data
distribution, but it is always the same as the training point
distribution.  Then the nearest neighbor search was performed on these
query points, and the results were averaged over all queries.
Although we ran a wide variety of experiments, for the sake of
conciseness we show only a few representative cases.  For all of the
experiments described here, we used 4000 data points in dimension 20 for
each data set, and there were 12,000 queries run for each data set.  For
the minimum-ambiguity method, the number of training points was 36,000.

The value of $\epsilon$ was either 1, 2, or 3 (allowing the reported
point to be a factor of 2, 3, or 4 further away than the true nearest
neighbor, respectively).  We computed the exact nearest neighbors
off-line to guage the algorithm's actual performance.  The reason for
allowing approximation errors is that in moderate to high dimensions,
the search times are typically smaller by orders of magnitude.  Also the
errors that were observed are typically quite a bit smaller on average
than these bounds (see Fig.~\ref{avgerr.tab}).  Note that average error
committed was typically only about $1/30$ of the allowable error.  The
maximum error was computed for each run of 12,000 query points, and then
averaged over all runs.  Even this maximum error was only around $1/4$
of the allowed error.  Some variation (on the order of a factor of 2)
was observed depending on the choice of search tree and point
distributions.

\begin{figure}
  \begin{center}
  \begin{tabular}{||c|c|c|c||}
    \hline\hline
 	$\epsilon$	& Avg. error	& Std. dev.	& Max. Error \\
    \hline
 	1.0		& 0.03643	& 0.0340	& 0.248 \\
	2.0		& 0.06070	& 0.0541	& 0.500 \\
	3.0		& 0.08422	& 0.0712	& 0.687 \\
    \hline\hline
  \end{tabular}
  \end{center}
  \caption{Average error commited, the standard deviation of the error,
  and the maximum error versus the allowed error, $\epsilon$.  Values
  were averaged over all runs.}
  \label{avgerr.tab}
\end{figure}

\subsection{Distributions Tested}

The distributions that were used in our experiments are listed below.
The clustered-gaussian distribution is designed to model point sets that
are clustered, but in which each cluster is full-dimensional.  The
clustered-orthogonal-ellipsoid and clustered-ellipsoid distributions are
both explicitly designed to model point distributions which are
clustered, and the clusters themselves are flat in the sense that the
points lie close to a lower dimensional subspace.  In the first case the
ellipsoids are aligned with the axes, and in the other case they are
more arbitrarily oriented.

\begin{description}
\item[Uniform:]
	Each coordinate was chosen uniformly from the interval $[-1,1]$.
\item[Clustered-gaussian:]
	The distribution is given a number of color classes $c$, and
	a standard deviation $\sigma$.  We generated $c$ points from
	the uniform distribution, which form cluster centers.  Each point
	is generated from a gaussian distribution centered at a randomly
	chosen cluster center with standard deviation $\sigma$.
\item[Clustered-orthogonal-ellipsoids:]
	The distribution can be viewed as a degenerate clustered-gaussian
	distribution where the standard deviation of each coordinate is
	chosen from one of two classes of distributions, one with a large
	standard deviation and the other with a small standard deviation.
	The distribution is specified by the number of color classes $c$
	and four additional parameters:
	\begin{itemize}
	\item	$\dmax$ is the maximum number of fat dimensions.
	\item	$\siglo$ and $\sighi$ are the minimum and maximum bounds
		on the large standard deviations, respectively (for the
		fat sides of the ellipsoid).
	\item	$\sigthin$ is the small standard deviation (for the thin
		sides of the ellipsoid).
	\end{itemize}
	Cluster centers are chosen as in the clustered-gaussian
	distribution.  For each color class, a random number $d'$
	between $1$ and $\dmax$ is generated, indicating the number of
	fat dimensions.  Then $d'$ dimensions are chosen at random to be
	fat dimensions of the ellipse.  For each fat dimension the
	standard deviation for this coordinate is chosen uniformly from
	$[\siglo,\sighi]$, and for each thin dimension the standard
	deviation is set to $\sigthin$.  The points are then generated
	by the same process as clustered-gaussian, but using these
	various standard deviations.
\item[Clustered-ellipsoids:]
	This distribution is the result of applying $d$ random rotation
	transformations to the points of each cluster about its center.
	Each cluster is rotated by a different set of rotations.  Each
	rotation is through a uniformly distributed angle in the range
	$[0,\pi/2]$ with respect to two randomly chosen dimensions.
\end{description}

In our experiments involving both clustered-orthogonal-ellipsoids and
clustered-ellipsoids, we set the number of clusters to 5, $\dmax = 10$,
$\siglo = \sighi = 0.3$, and $\sigthin$ varied from $0.03$ to $0.3$.
Thus, for low values of $\sigthin$ the ellipsoids are relatively flat,
and for high values this becomes equivalent to a clustered-gaussian
distribution with standard deviation of 0.3.

\subsection{Data and Query Points from the Same Distribution}

\begin{figure}[htb]
  \begin{center}
  \begin{tabular}{cc}
  \psfig{figure=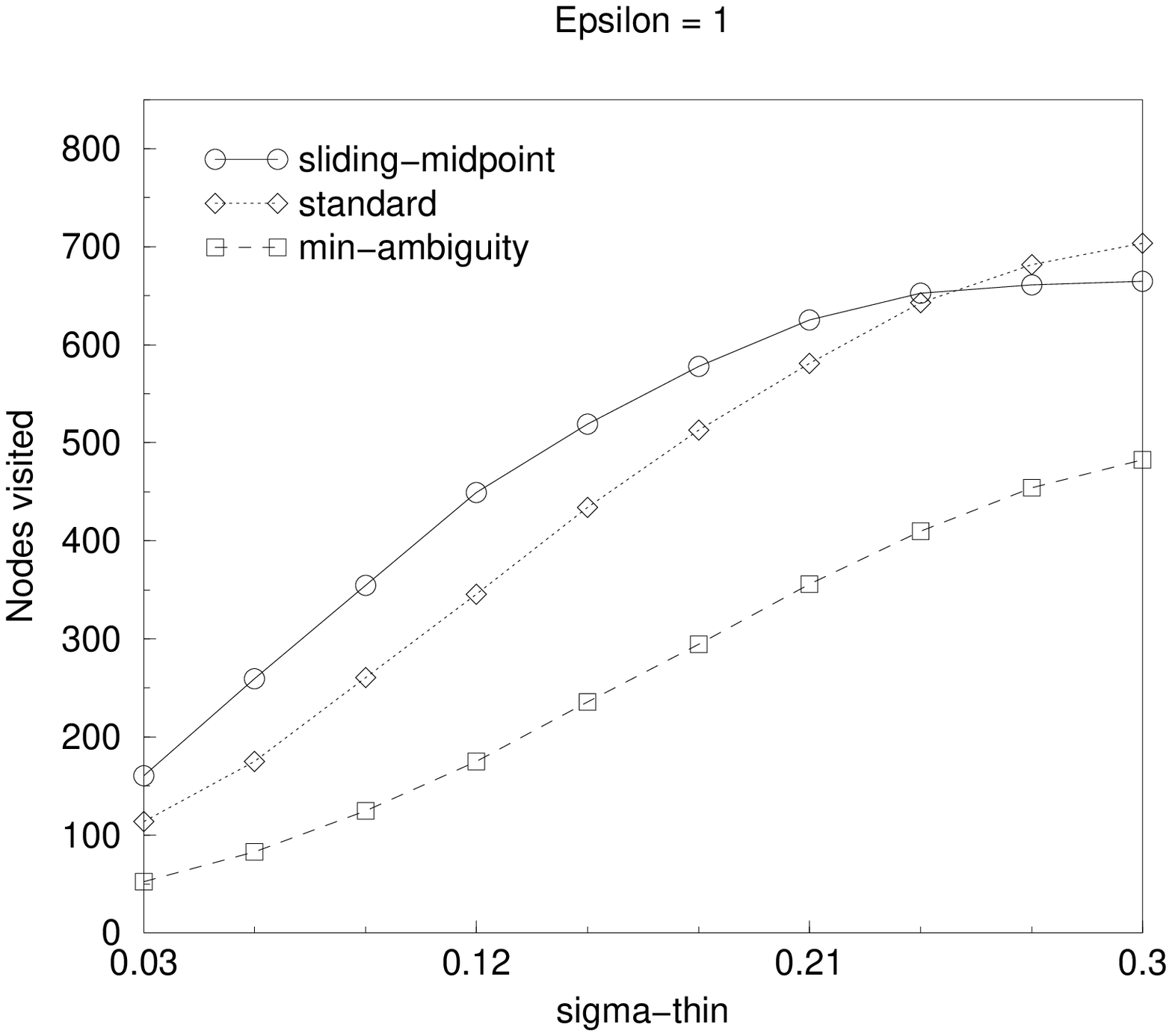,width=2.36in} &
  \psfig{figure=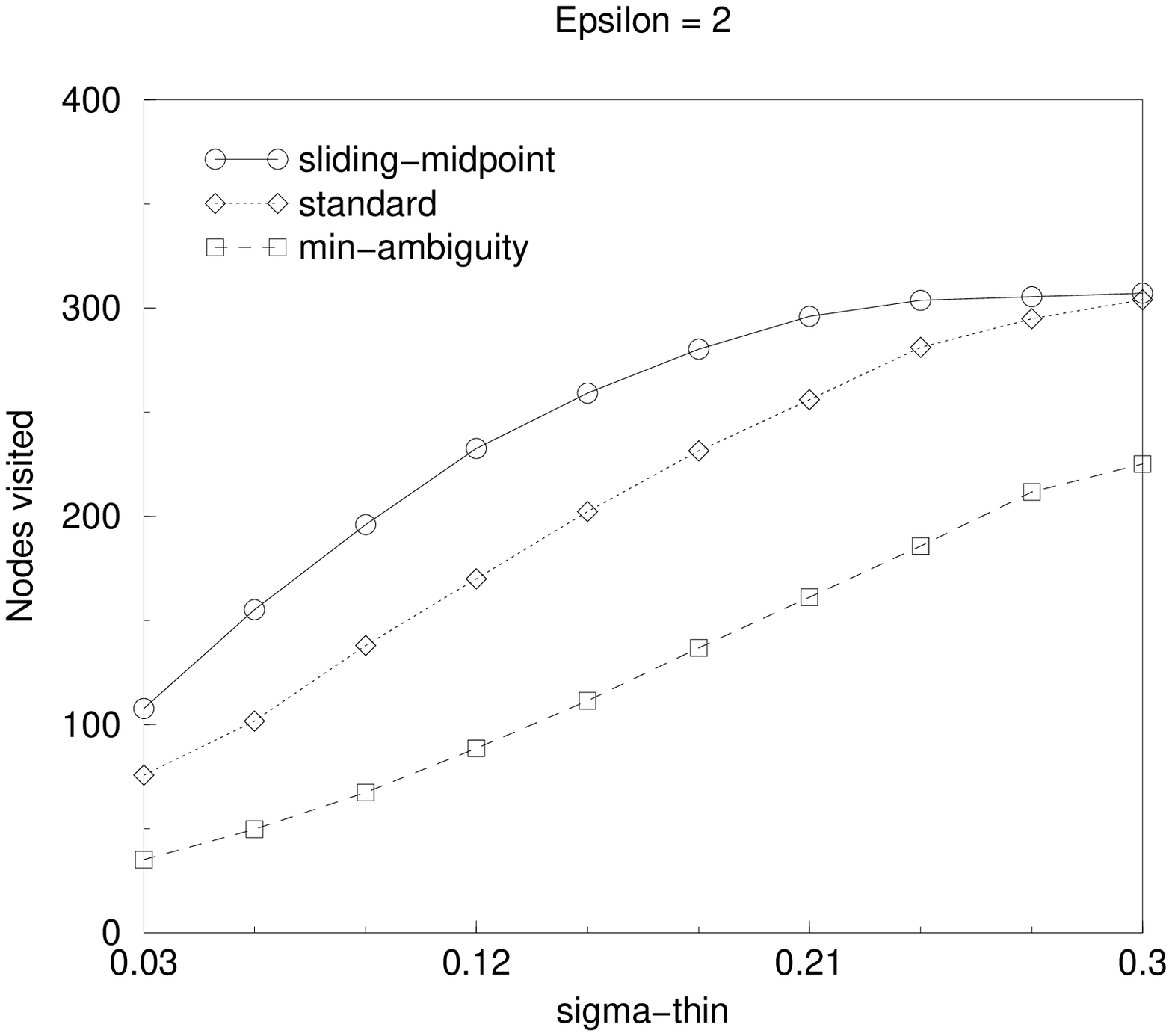,width=2.36in} \\
  (a) & (b)
  \end{tabular} \\
  \centerline{\psfig{figure=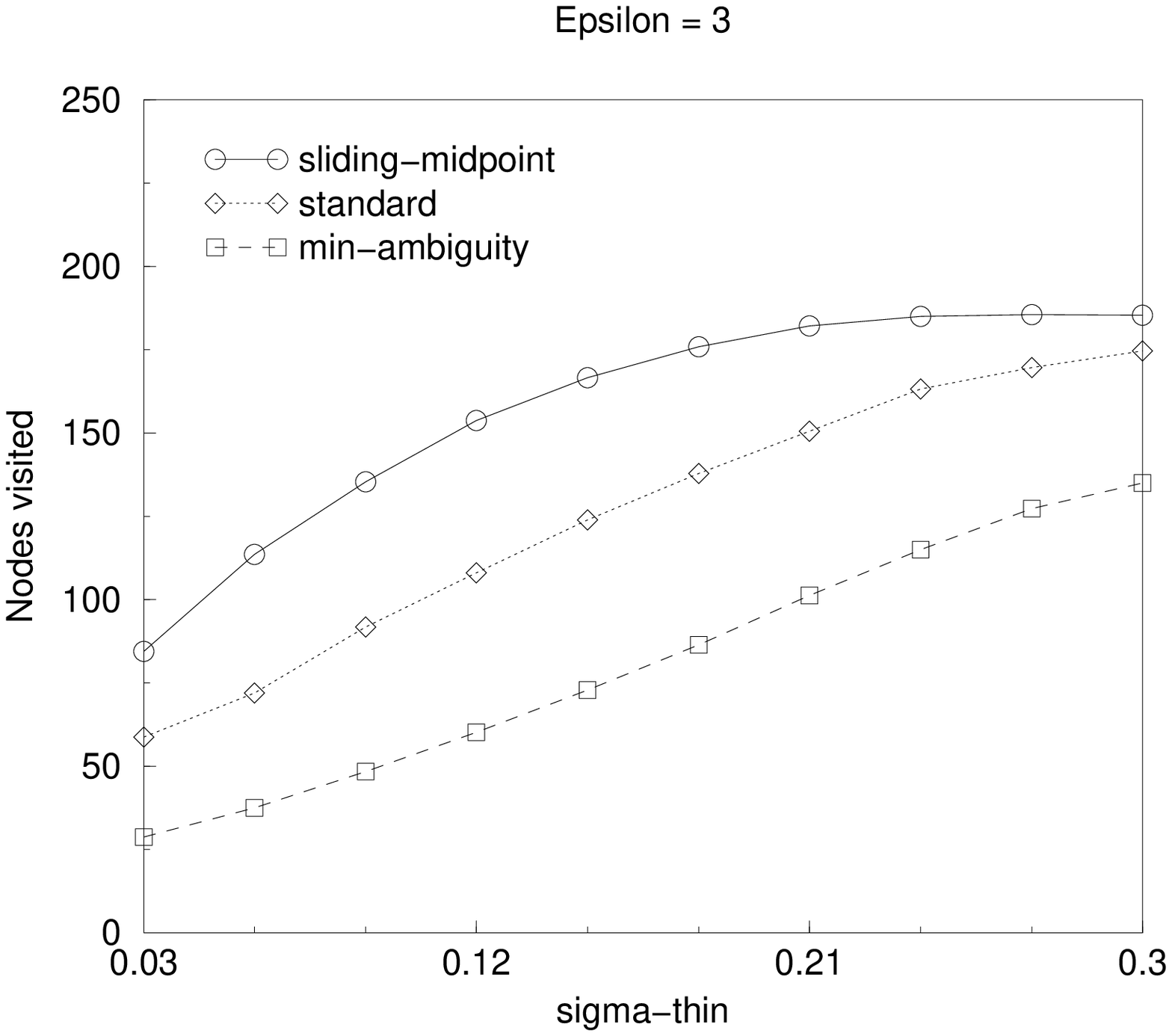,width=2.36in}}
  (c)
  \end{center}
  \caption{Number of nodes visited versus $\sigthin$ for
  	$\epsilon \in \{0, 1, 2\}$.
  	Data and query points both sampled from the same
  	clustered-orthogonal-ellipsoid distribution.}
  \label{dce-qce.fig}
\end{figure}

\begin{figure}[htb]
  \begin{center}
  \begin{tabular}{cc}
  \psfig{figure=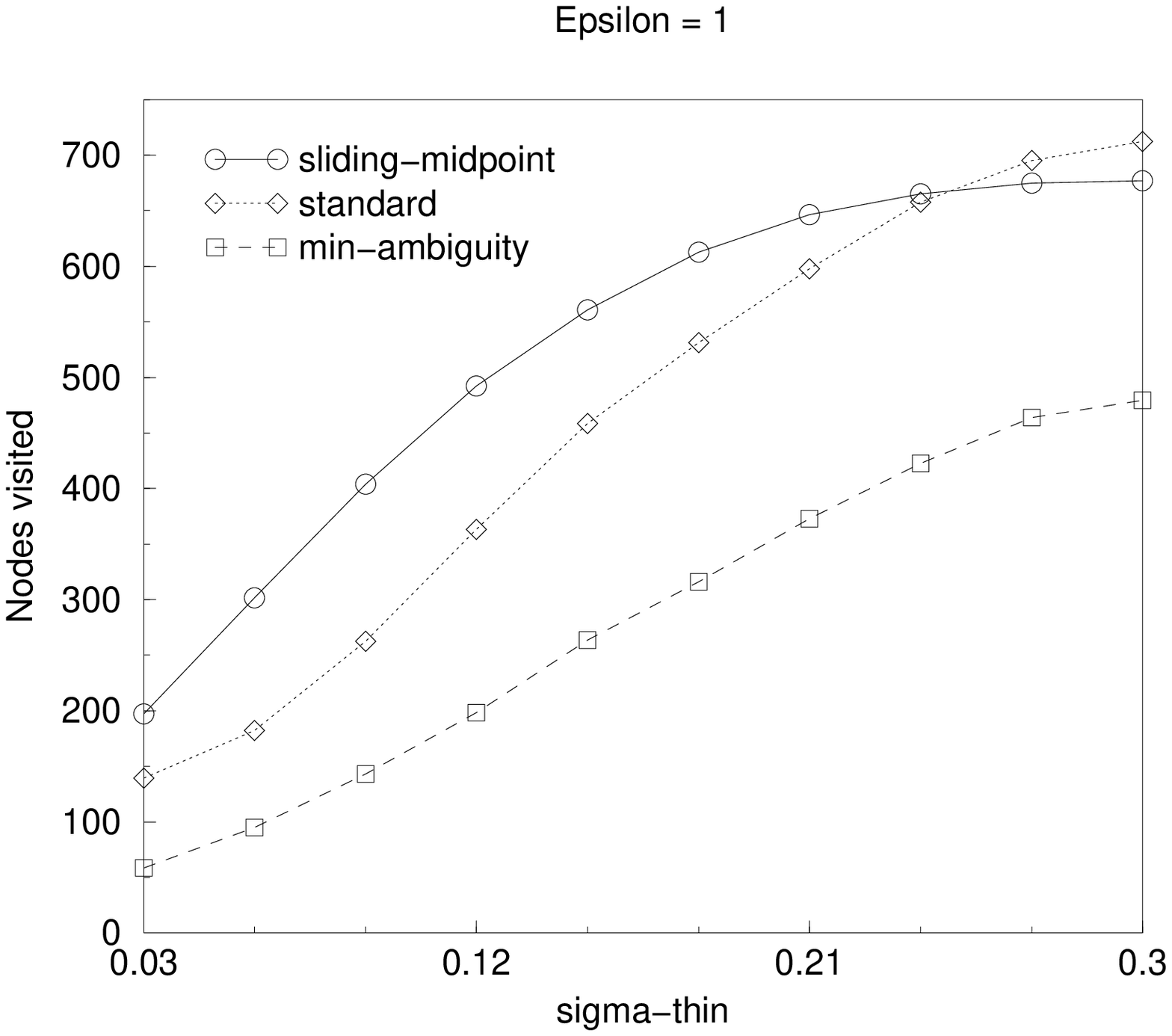,width=2.36in} &
  \psfig{figure=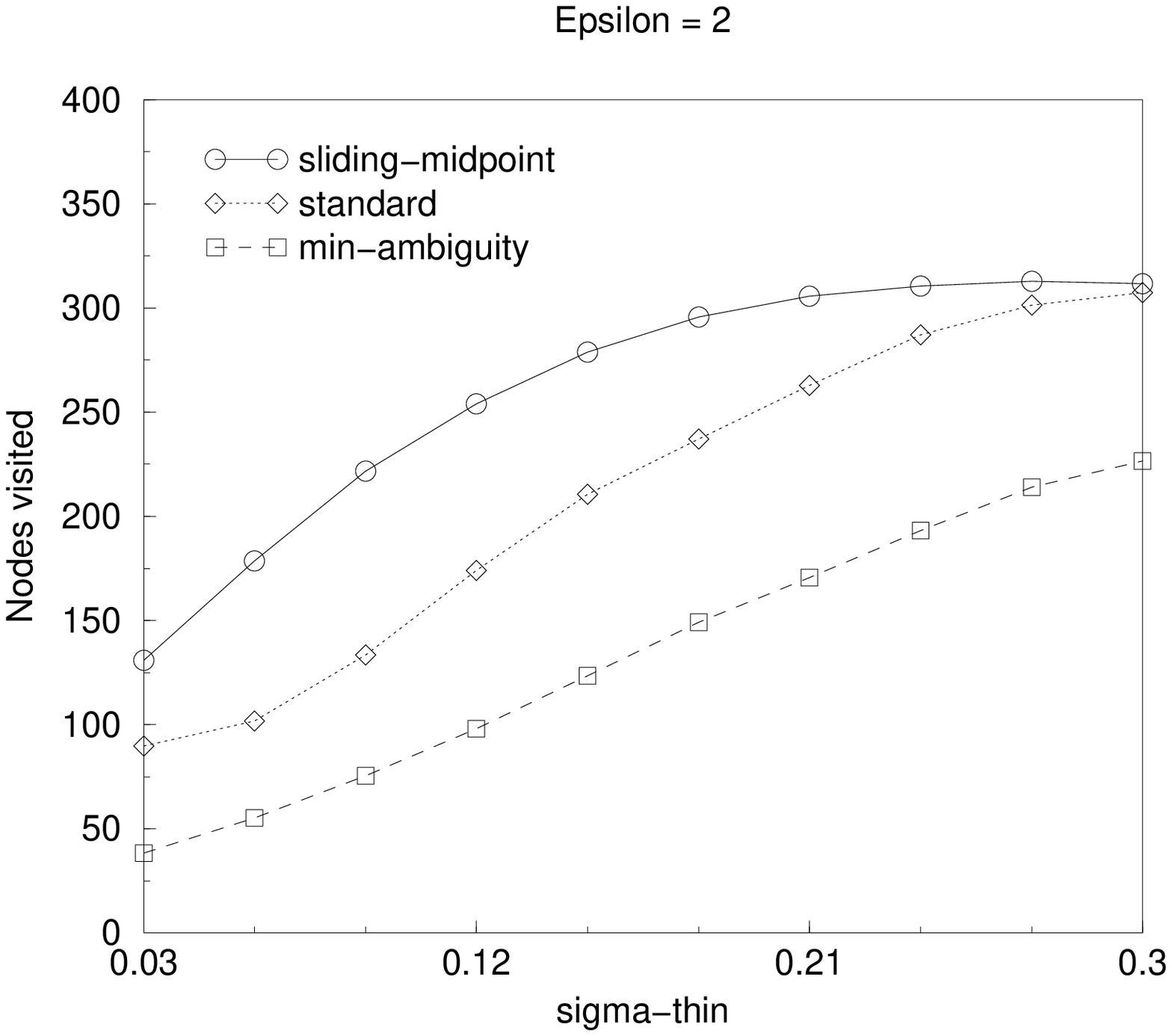,width=2.36in} \\
  (a) & (b)
  \end{tabular} \\
  \centerline{\psfig{figure=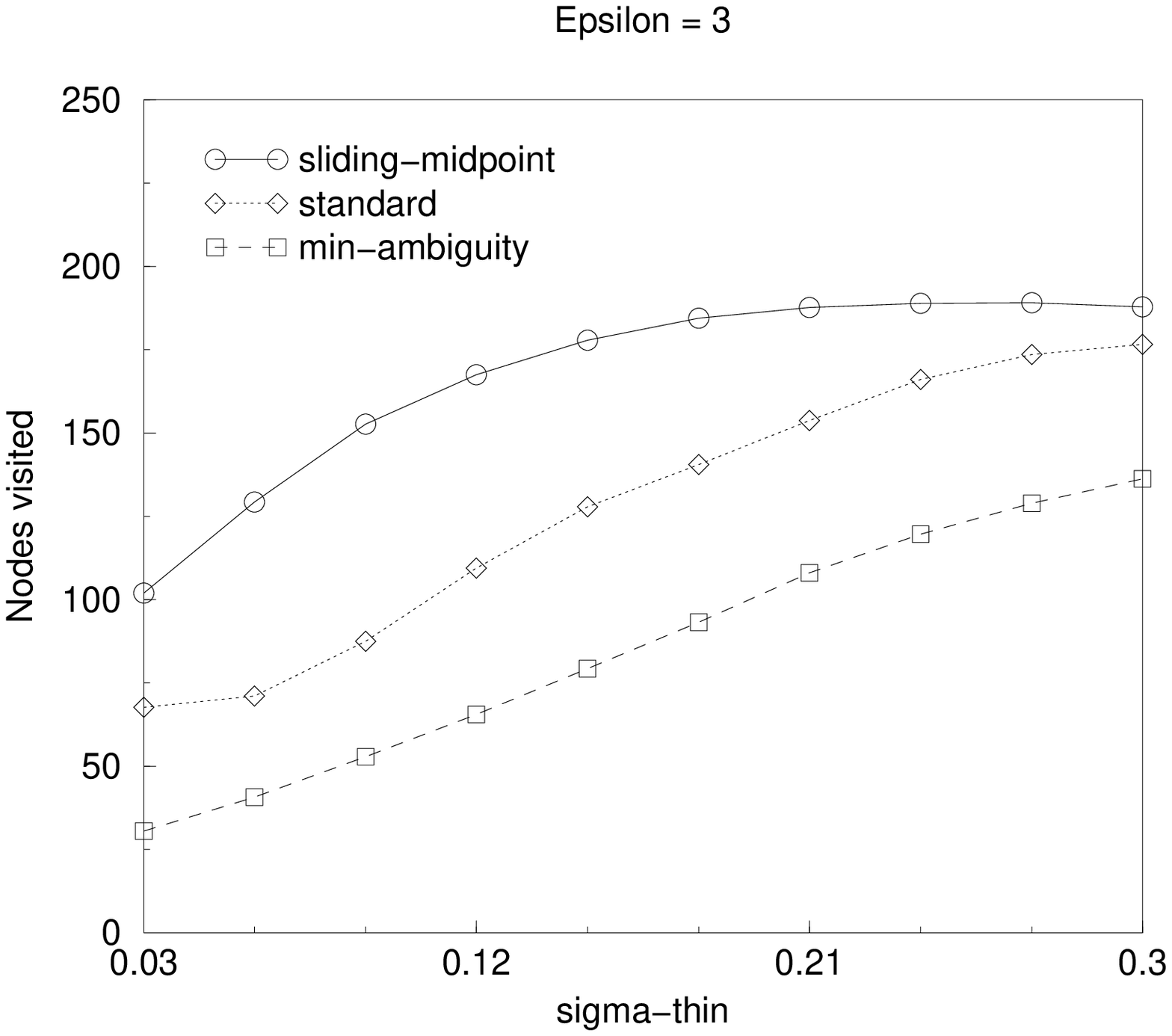,width=2.36in}}
  (c)
  \end{center}
  \caption{Number of nodes visited versus $\sigthin$ for
  	$\epsilon \in \{0, 1, 2\}$.
  	Data and query points both sampled from the same
  	clustered-ellipsoid distribution.}
  \label{dcr-qcr.fig}
\end{figure}

For our first set of experiments, we considered data and query points
from the same clustered distributions.  We considered both
clustered-orthogonal-ellipsoids and clustered-ellipsoid distributions in
Figs.~\ref{dce-qce.fig} and~\ref{dcr-qcr.fig}, respectively.  The three
different graphs are for (a) $\epsilon = 1$, (b) $\epsilon = 2$, and (c)
$\epsilon = 3$.  In all three cases the same clusters centers were
used.  Note that the graphs do not share the same $y$-range,
and in particular the search algorithm performs significantly faster
as $\epsilon$ increases.

Observe that all of the splitting methods perform better when $\sigthin$
is small, indicating that to some extent they exploit the fact that the
data points are clustered in lower dimensional subspaces.  The relative
differences in running time were most noticeable for small values of
$\sigthin$, and tended to diminish for larger values.

Although the minimum-ambiguity splitting method was designed for dealing
with data and query points from different distributions, we were
somewhat surprised that it actually performed the best of the three
methods in these cases.  For small values of $\sigthin$ (when
low-dimensional clustering is strongest) its average running time
(measured as the number of noded visited in the tree) was typically from
30-50\% lower than the standard splitting method, and over 50\% lower than
the sliding-midpoint method.  The standard splitting method typically
performed better than the sliding-midpoint method, but the difference
decreased to being insignificant (and sometimes a bit worse) as
$\sigthin$ increased.

\subsection{Data and Query Points from Different Distributions}

\begin{figure}[htb]
  \begin{center}
  \begin{tabular}{cc}
  \psfig{figure=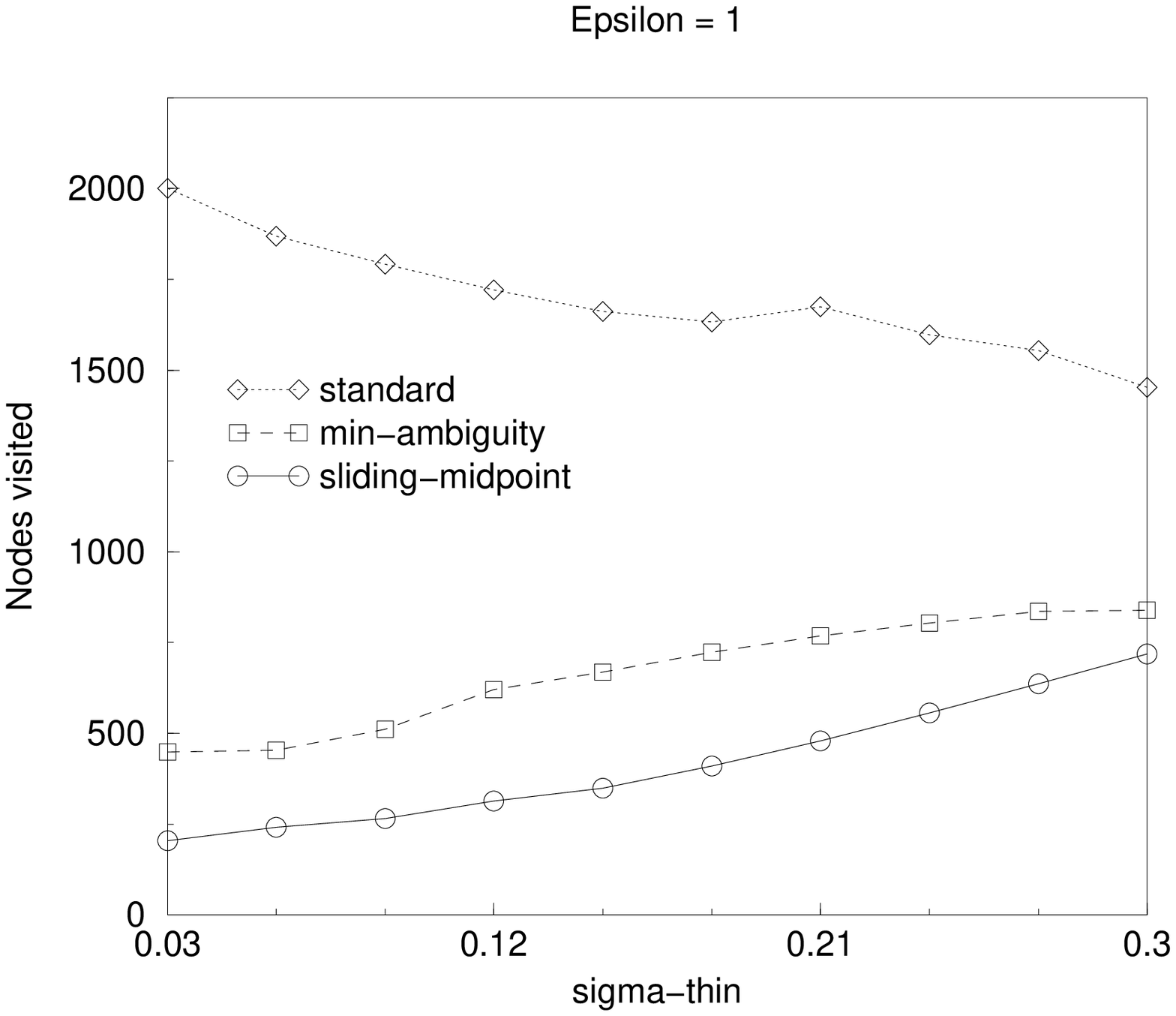,width=2.36in} &
  \psfig{figure=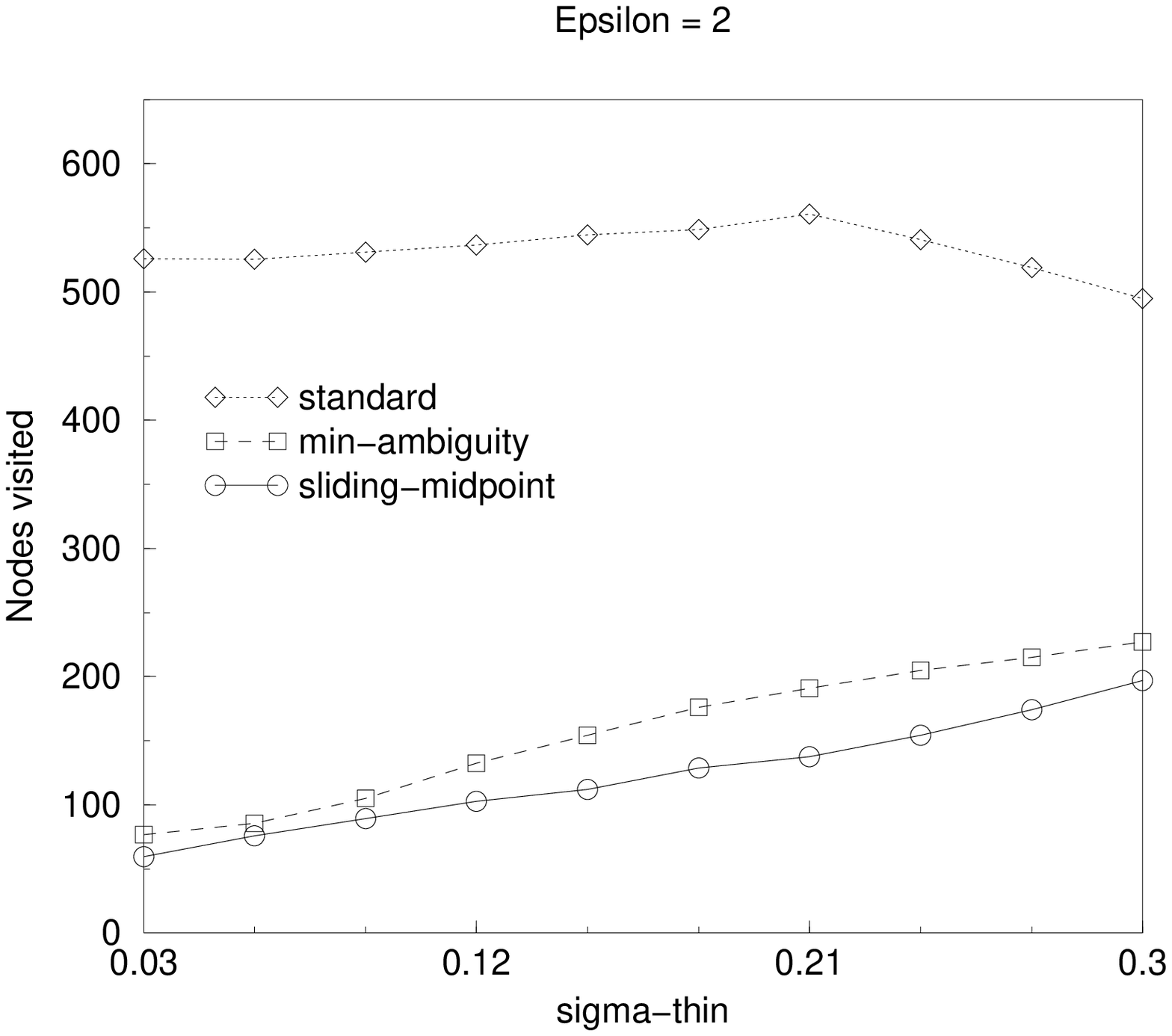,width=2.36in} \\
  (a) & (b)
  \end{tabular} \\
  \centerline{\psfig{figure=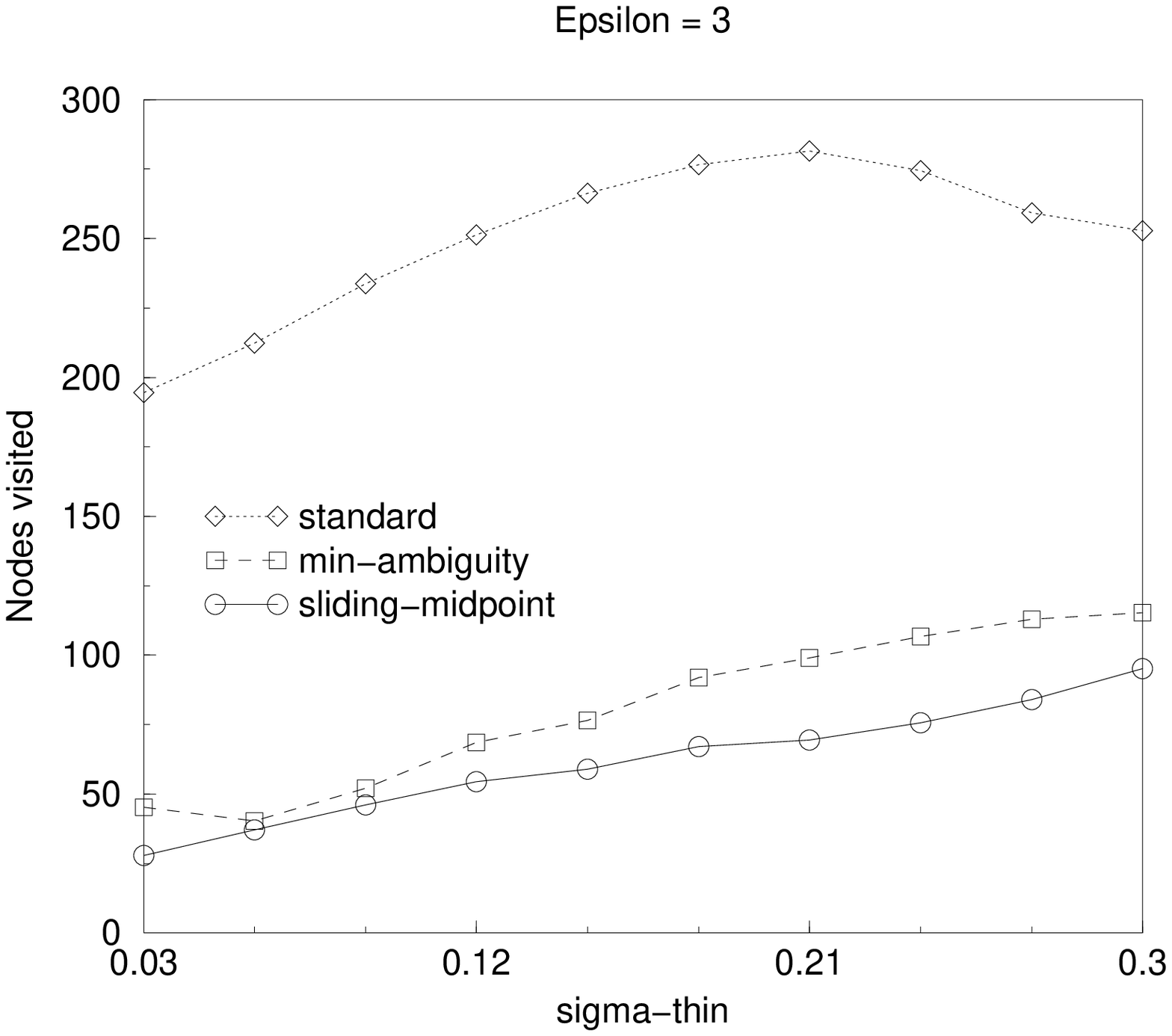,width=2.36in}}
  (c)
  \end{center}
  \caption{Number of nodes visited versus $\sigthin$ for
  	$\epsilon \in \{0, 1, 2\}$.
  	Data sampled from the clustered-orthogonal-ellipsoid distribution
  	and query points from the uniform distribution.}
  \label{dce-qu.fig}
\end{figure}

\begin{figure}[htb]
  \begin{center}
  \begin{tabular}{cc}
  \psfig{figure=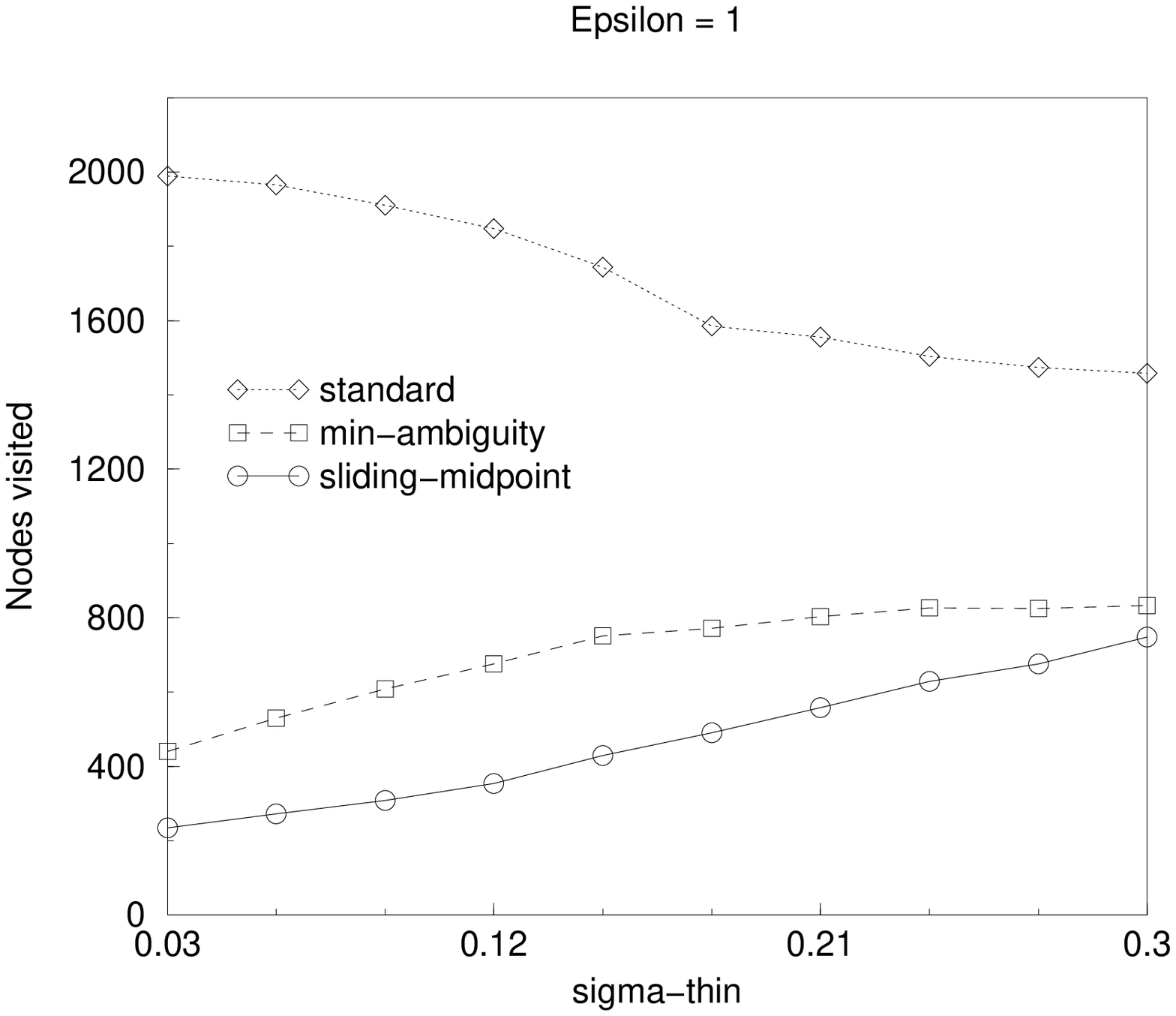,width=2.36in} &
  \psfig{figure=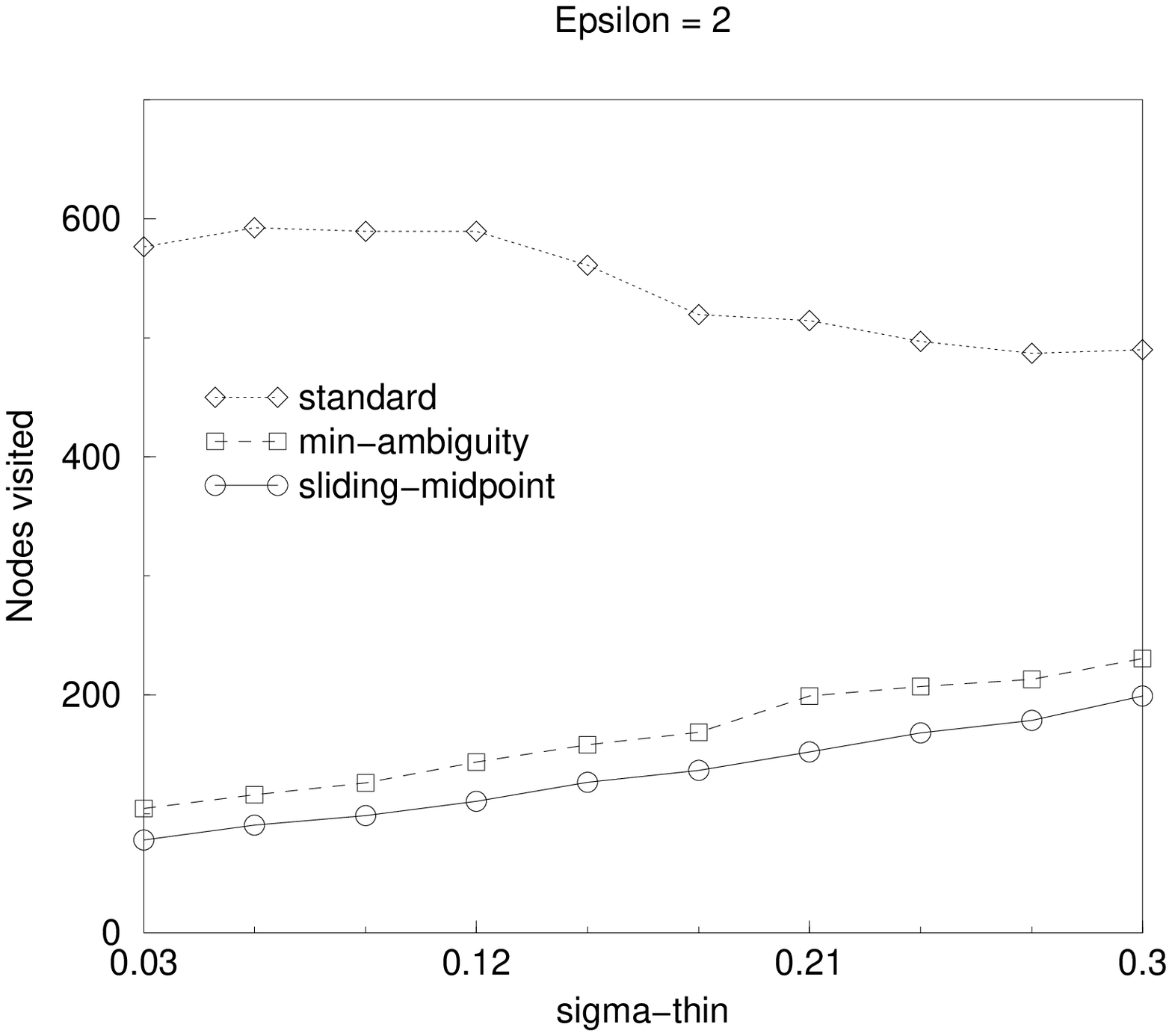,width=2.36in} \\
  (a) & (b)
  \end{tabular} \\
  \centerline{\psfig{figure=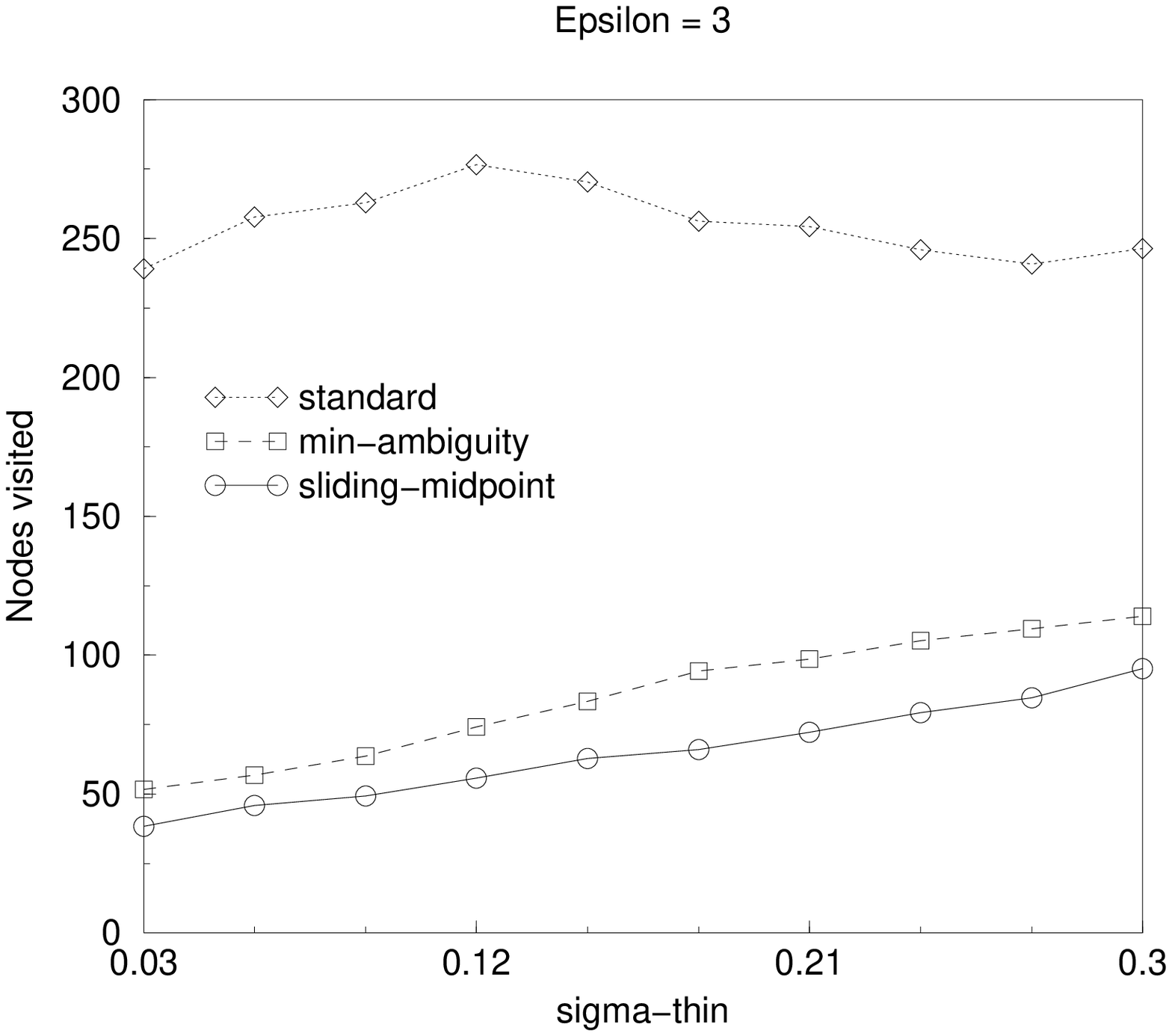,width=2.36in}}
  (c)
  \end{center}
  \caption{Number of nodes visited versus $\sigthin$ for
  	$\epsilon \in \{0, 1, 2\}$.
  	Data sampled from the clustered-ellipsoid distribution
  	and query points from the uniform distribution.}
  \label{dcr-qu.fig}
\end{figure}

For our second set of experiments, we considered data points from a
clustered distribution and query points from a uniform distribution.
This particular choice was motivated by the situation shown in
Fig.~\ref{slmid.fig}, where the standard splitting method can produce
cells with high aspect ratios.  For the data points we considered both
the clustered-orthogonal-ellipsoids and clustered-ellipsoid
distributions in Figs.~\ref{dce-qu.fig} and~\ref{dcr-qu.fig},
respectively.  As before, the three different graphs are for (a)
$\epsilon = 1$, (b) $\epsilon = 2$, and (c) $\epsilon = 3$.  Again,
note that the graphs do not share the same $y$-range.

Unlike the previous experiment, overall running times did not vary
greatly with $\sigthin$.  Sometimes running times increased moderately
and other times they decreased moderately as a function of $\sigthin$.
However, there were significant differences between the standard
splitting method, which consistently performed much worse than the other
two methods.  For the smallest values of $\sigthin$, there was around a
5-to-1 difference in running time between then standard method and
sliding-midpoint.

For larger values of $\epsilon$ (2 and 3) the performance of
sliding-midpoint and minimum-ambiguity were very similar, with
sliding-midpoint having the slight edge.  It may seem somewhat
surprising that minimum-ambiguity performed significantly worse (a
factor of 2 to 3 times worse) than sliding-midpoint, since
minimum-ambiguity was designed exactly for this the situation where
there is a difference between data and query distributions.  This may be
due to limitations on the heuristic itself, or the limited size of the
training set.  However, it should be kept in mind that sliding-midpoint
was specially designed to produce large empty cells in the uncluttered
regions outside the clusters (recall Fig.~\ref{slmid.fig}).

\subsection{Construction Times}

The results of the previous sections suggest that the minimum ambiguity
splitting produces trees that can answer queries efficiently for a
variety of point and data distributions.  Its main drawback is the
amount of time that it takes to build the tree.  Both the standard and
sliding-midpoint methods can be built quite efficiently in time $O(nh)$,
where $n$ is the number of data points, and $h$ is the height of the
tree.  The standard kd-tree has $O(\log n)$ height, and while the
sliding-midpoint tree need not have $O(\log n)$ height, this seems to be
true for many point distributions.  For the 4000 point data sets in
dimension 20, both of these trees could be constructed in under 10 CPU
seconds.

However, the construction time for the minimum-ambiguity tree is quite a
bit higher.  It can be argued that the time to construct the tree is
roughly (within logarithmic factors) proportional to the time to compute
the (approximate) nearest neighbors for all the training points.  In
order to construct the tree, first the nearest neighbors for each of the
training points must be computed.  This is done in an auxiliary nearest
neighbor tree, e.g., one built using the standard or sliding-midpoint
method.  Then to determine the splitting hyperplane for each cell of the
minimum-ambiguity tree, requires consideration of all the nearest
neighbor balls that overlap the current cell.  However, in order to
compute the nearest neighbors of the training points, each point whose
nearest neighbor ball overlaps the cell would have to visit the cell in
any case.

\begin{figure}[htb]
  \begin{center}
  \begin{tabular}{cc}
  \psfig{figure=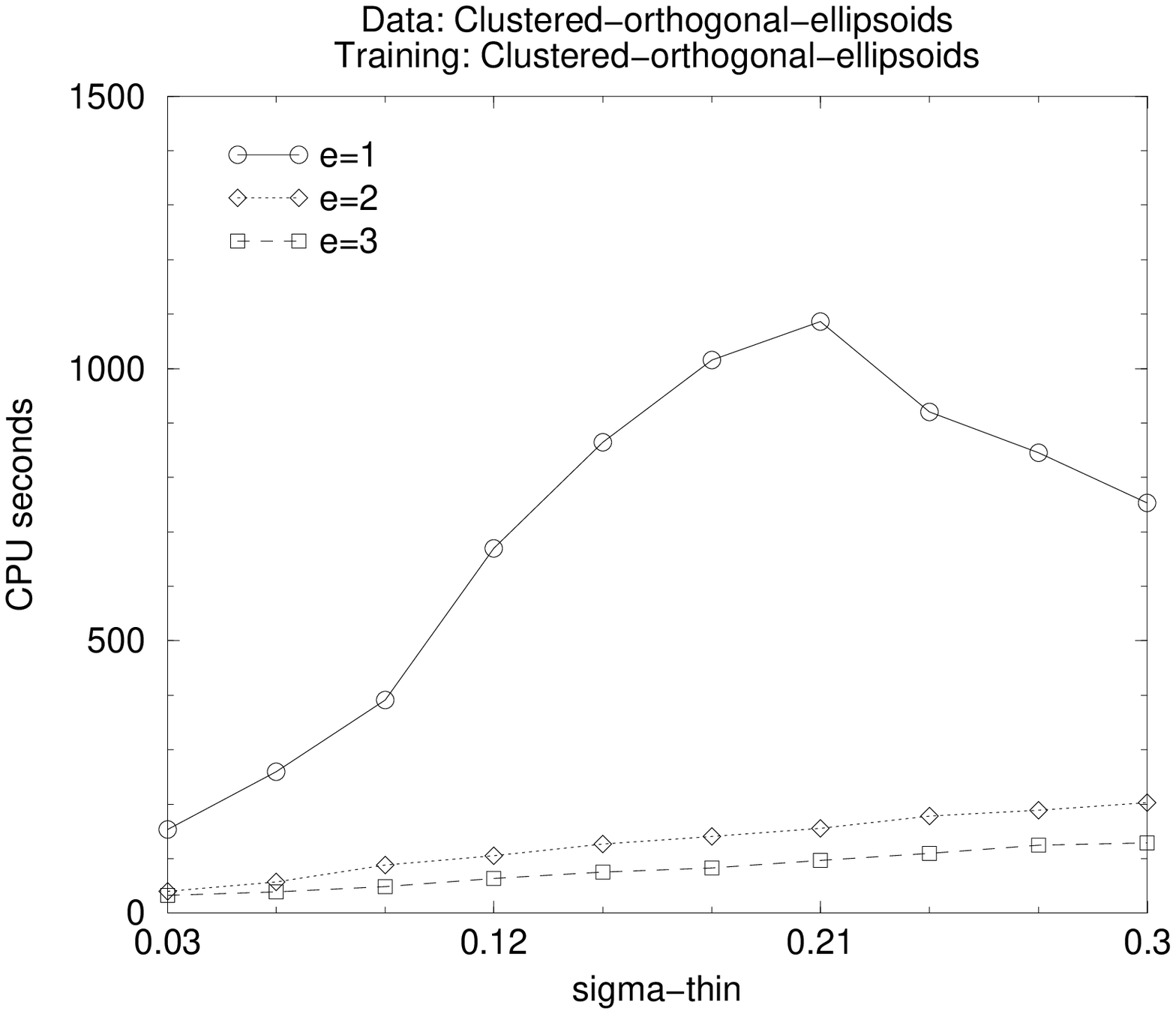,width=2.36in} &
  \psfig{figure=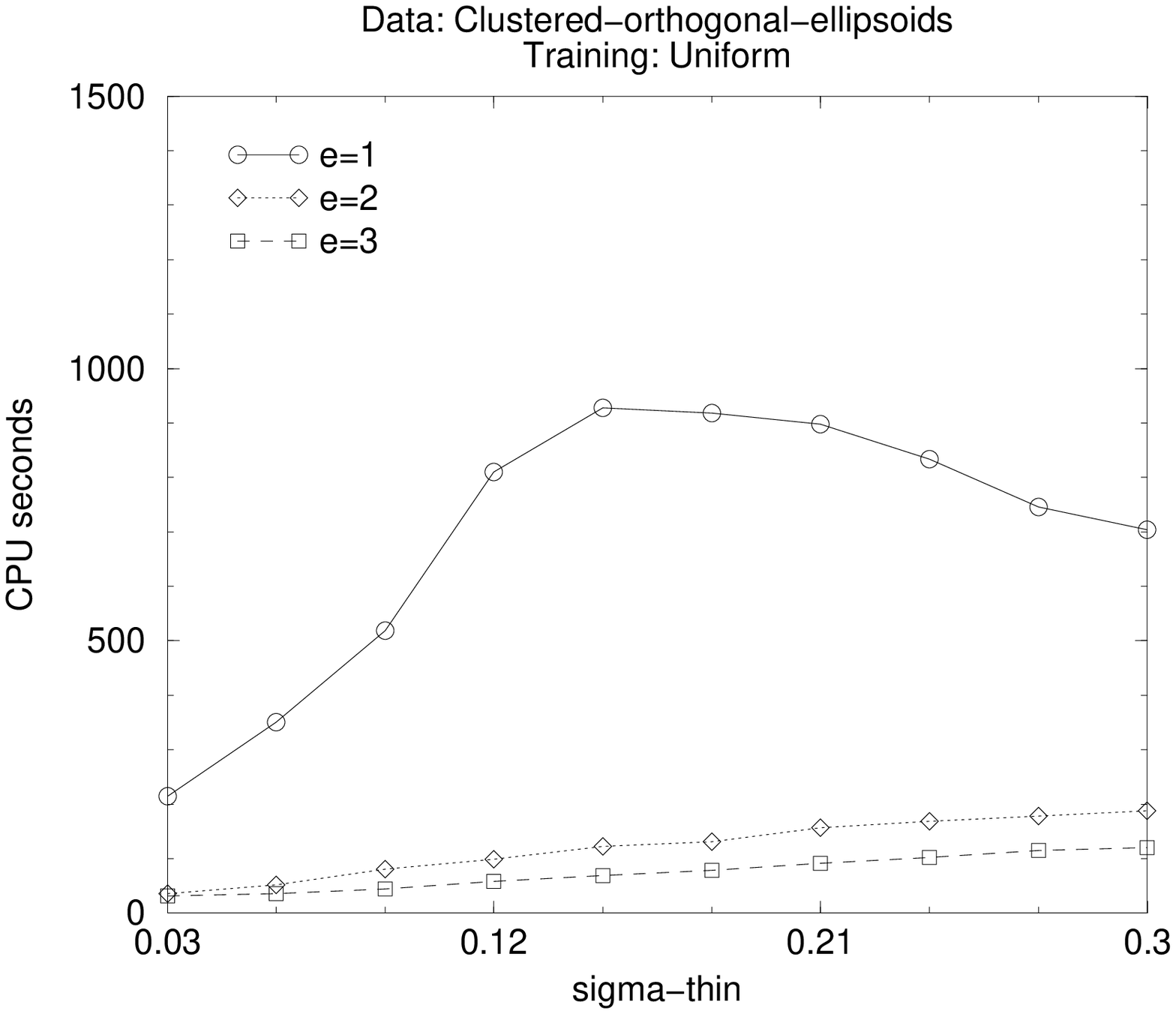,width=2.36in} \\
  (a) & (b) \\
  ~ \\
  \psfig{figure=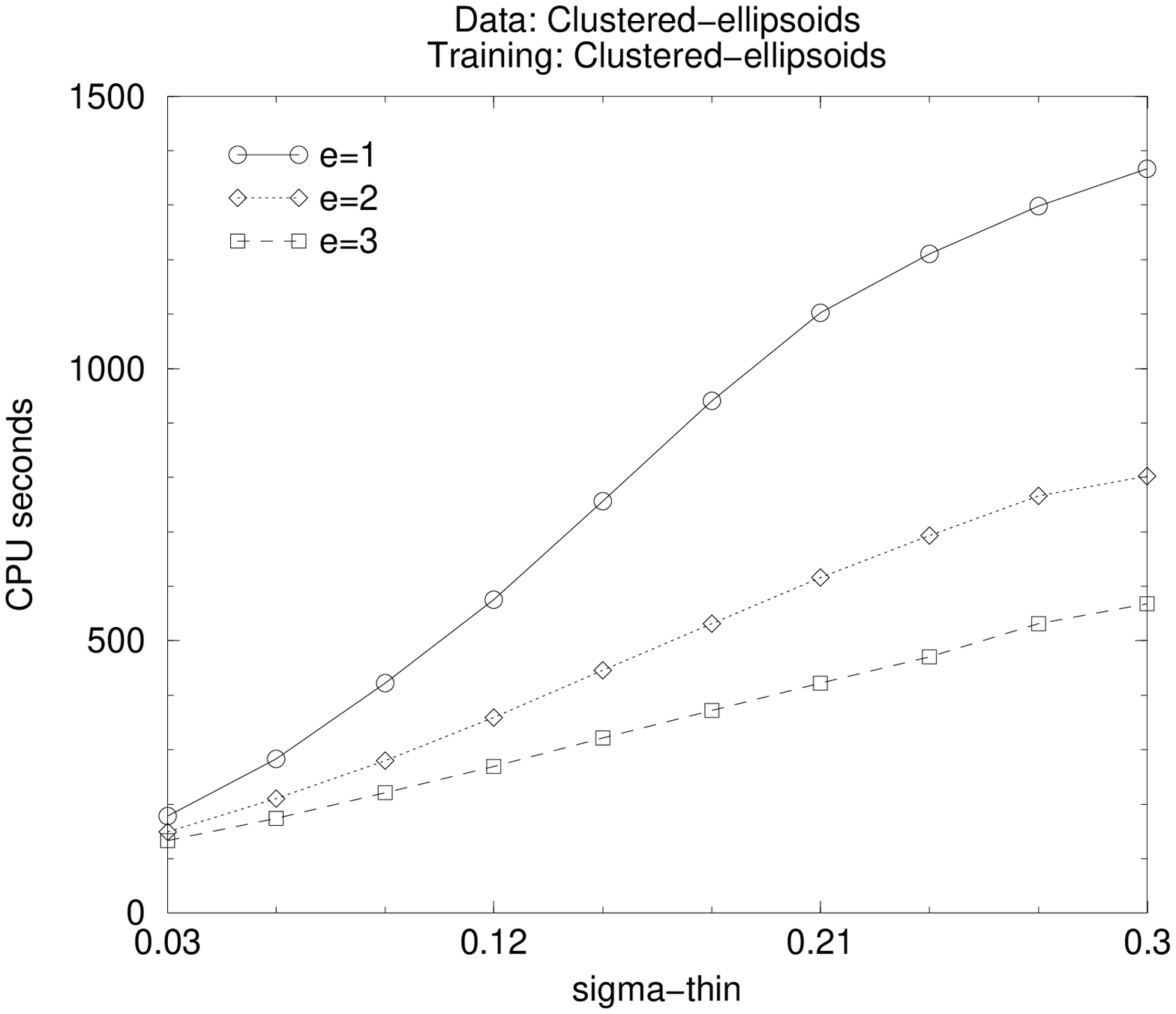,width=2.36in} &
  \psfig{figure=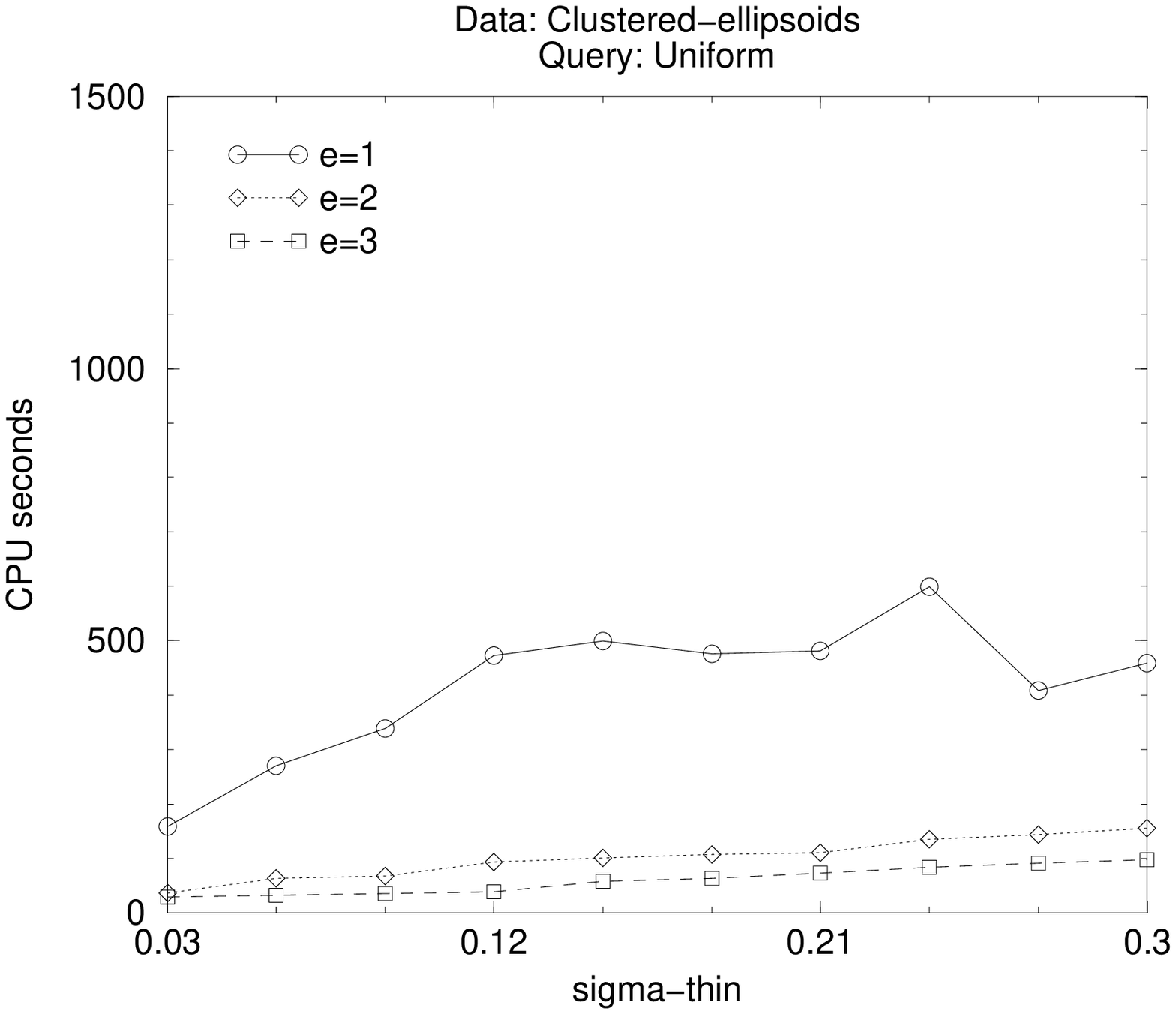,width=2.36in} \\
  (c) & (d)
  \end{tabular} 
  \end{center}
  \caption{Time to construct minimum-ambiguity tree versus
	$\sigthin$.}
  \label{ma-const.fig}
\end{figure}

Since we used 9 times the number of data points as training points, it
is easy to see that the minimum-ambiguity tree will take much longer
to compute than the other two trees.  Notice that when $\epsilon > 0$,
we compute nearest neighbors approximately, and so this can offer an
improvement in construction time.  In Fig.~\ref{ma-const.fig} we
present the construction time for the minimum-ambiguity tree for 
various combinations of data and training distributions.  Observe
that the construction times are considerably greater than those for
the other two methods (which were under 10 CPU seconds), and that
the construction time is significantly faster for higher values
of $\epsilon$.

\section{Conclusions}

In this paper we have presented an empirical analysis of two new
splitting methods for kd-trees: sliding-midpoint and minimum-ambiguity.
Both of these methods were designed to remedy some of the deficiencies
of the standard kd-tree splitting method, with respect to data
distributions that are highly clustered in low-dimensional subspaces.
Both methods were shown to be considerably faster than the standard
splitting method in answering queries when data points were drawn from a
clustered distribution and query points were drawn from a uniform
distribution.  The minimum-ambiguity method performed better when both
data and query points were drawn from a clustered distribution. But this
method has a considerably higher construction time.  The
sliding-midpoint method, while easy to build, seems to perform sometimes
better and sometimes worse than the standard kd-tree splitting method.

The enhanced performance of the minimum-ambiguity method suggests that
even within the realm of kd-trees, there may be significant improvements
to be made by fine-tuning the structure of the tree to the data and
query distributions.  However, because of its high construction cost, it
would be nice to determine whether there are other heuristics that would
lead to faster construction times.  This suggest the intriguing
possibility of search trees whose structure adapts dynamically to the
structure of queries over time.  The sliding-midpoint method raises hope
that it may be possible to devise a simple and efficiently computable
splitting method, that performs well across a wider variety of
distributions than the standard splitting method.

\section{Acknowledgements}
We would like to thank Sunil Arya for helpful discussions on the
performance of the sliding-midpoint method.



\end{document}